\documentclass[prl,aps,twocolumn,groupedaddress,floats,showpacs,final]{revtex4}
\usepackage{graphicx}
\usepackage{dcolumn}
\usepackage{bm}
\usepackage{color}
\usepackage{ulem}
\definecolor{blue}{rgb}{0.3,0.3,0.9}

\def\JQ{$J$-$Q$ }
\def\beq{\begin{eqnarray}}
\def\eeq{\end{eqnarray}}
\def\i{{\rm i}}

\begin{document}

\author{Kun Chen$^{1,2}$}
\author{Yuan Huang$^{1,2}$}
\author{Youjin Deng$^{1,2}$}
\email{yjdeng@ustc.eud.cn}
\author{A.B. Kuklov$^3$}
\email{Anatoly.Kuklov@csi.cuny.edu}
\author{N.V. Prokof'ev$^{2,4}$}
\email{prokofev@physics.umass.edu}
\author{B.V. Svistunov$^{2,4}$}
\email{svistunov@physics.umass.edu}

\affiliation{$^1$Hefei National Laboratory for Physical Sciences at Microscale and Department of Modern Physics,
University of Science and Technology of China, Hefei, Anhui 230026, China}
\affiliation{$^2$Department of Physics, University of Massachusetts, Amherst, Massachusetts 01003, USA}
\affiliation{$^3$Department of Engineering Science and Physics,
CSI, CUNY, Staten Island, New York 10314, USA}
\affiliation{$^4$Russian Research Center ``Kurchatov Institute'',123182 Moscow, Russia}

\title{Deconfined Criticality Flow in the Heisenberg Model with Ring-Exchange 
Interactions}

\date{\today}
\begin{abstract}
Quantum transition points in the \JQ model -- the test bed of the deconfined 
critical point theory -- and the SU(2)-symmetric discrete noncompact CP$^1$ representation 
of the deconfined critical action are directly compared by the flowgram method.
We find that the flows of two systems coincide in a broad region
of linear system sizes ($10<L<50$  for the \JQ model), implying that the 
deconfined critical point theory correctly captures the mesoscopic physics of competition between the antiferromagnetic 
and valence-bond orders in quantum spin systems.
At larger sizes, however, we observe significant deviations between the two flows 
which both demonstrate strong violations of scale invariance. 
This reliably rules out the second-order transition scenario in at least one of the two models 
and suggests the most likely explanation for the nature of the transition in the \JQ model.
\end{abstract}

\pacs{64.70.Tg, 05.30.-d, 05.50.+q, 75.10.-b}

\maketitle

The concept of the deconfined critical point (DCP) \cite{Motrunich,dcp,dcp1} was developed for understanding 
quantum transitions in two dimensions (2D) between phases characterized by different broken symmetries.
The key feature of DCP is the emergence of fractional degrees of freedom (spinons) and gauge fields
at the critical point (cf. \cite{Sondhi}). Potentially, the DCP scenario has a broad range of applications
ranging from quantum phases transitions in lattice models and magnets to normal-superfluid transitions
in multicomponent charged superconductors, etc. \cite{Motrunich,dcp,dcp1,DCPbabaev,Sachdev2009}. 
Ultracold atoms in an optical lattice is another promising system where DCP can be tested experimentally \cite{dcpatom}.

A hallmark of the theory is a conjecture that the DCP universality class is captured by the 
3D classical DCP action involving two complex-valued matter fields, $\psi_{a=1,2}$, describing spinons
coupled to a vector gauge field \cite{Motrunich,dcp,dcp1,Sachdev2009}.
Depending on the symmetry group of the underlying quantum system---global U(1) or global SU(2)---the DCP action
features the following symmetry in terms of its two components: either the Z$_2$ symmetry between 
two spinon fields and the U(1)$\times$U(1) symmetry associated with the individual phases of $\psi_a$ or an enhanced  
SU(2) symmetry between the spinon fields. However, flowgram studies of the typical U(1)$\times$U(1) \cite{flowgram}
and SU(2) \cite{dejavu} DCP actions revealed generic runaway flows consistent with weak first-order transitions for any value of the gauge interaction (cf. Refs.~\cite{anatoly2004,Sudbo2006}, where the first order was observed, respectively, in a special model, or at a specific value of the interaction).

The initial work, focused on microscopic models of the superfluid to solid quantum phase transitions,  
first claimed the observation of the second-order U(1)$\times$U(1) transition \cite{Sandvik2002}, but 
severe violations of scale invariance revealed in the subsequent analysis all but ruled it out \cite{Sandvik2006}.
Similarly to the U(1)$\times$U(1) case, early studies of the antiferromagnetic SU(2)-symmetric \JQ model \cite{Sandvik2007,Sandvik2009,Melko2008} suggested that the N\'eel phase transforms into the valence-bond solid 
(VBS) in a  continuous fashion, while subsequent work \cite{Wiese2008,Sandvik2010} revealed 
violations of scale invariance. It is important, however, that, up to linear system sizes of a few hundred sites, the \JQ
model clearly demonstrates an emergent $U(1)$ symmetry and its runaway flow remains rather weak, 
leaving room for speculations about the second-order DCP scenario \cite{Sandvik2010}.

In this Letter, we perform a direct quantitative comparison of critical flows in the \JQ 
and the 3D SU(2)-symmetric discrete noncompact CP$^1$ models. The rationale behind our study is as follows. 
Slow runaway flows in both models suggest the key point that, independently of the order of the transition, 
the DCP theory in general, and the 3D SU(2)-symmetric discrete noncompact CP$^1$ model, in particular, capture the essence 
of the quantum phase transition {\it at least} at intermediate scales of distances.  
And we indeed find that the winding-number flowgrams \cite{flowgram,dejavu} of the two models can be 
collapsed in a significantly large region of linear system sizes (up to $L \approx 75$  for the \JQ model), proving the hypothesis. 
At larger sizes we observe significant deviations between the two flows which preserve their 
runaway character. The most conservative conclusion, then, is that at least one 
of the two models does not feature the second-order criticality, with  the straightforward interpretation
being that both models feature weak first-order transitions.

\noindent {\it \JQ and DCP models}.
The SU(2)-symmetric \JQ model describing $s=1/2$ spins on a square lattice has been analyzed in Ref.~\cite{Sandvik2007}:
\beq
H = J\sum_{\langle ij\rangle}\hat{\vec{S}}_i  \hat{\vec{S}}_j -
Q\sum_{\langle ijkl\rangle} (\hat{\vec{S}}_i \hat{\vec{S}}_j-\hbox{$\frac{1}{4}$})
(\hat{\vec{S}}_k \hat{\vec{S}}_l-\hbox{$\frac{1}{4}$}) \;.
\label{JQ}
\eeq
The first sum runs over nearest-neighbor sites $\langle ij\rangle$ and represents the standard Heisenberg model with $J>0$. The second sum runs over the corners of plaquettes 
${\cal P}=\langle ijkl\rangle$ such that $ij$ and $kl$ form two parallel adjacent horizontal or
vertical links and add the four-site ring-exchange terms with $Q>0$.  
The model features a phase transition between the N\'eel and an insulating VBS phase 
(its precise nature, however, cannot be determined from available system sizes \cite{Wiese2008,Sandvik2012}).
While in the N\'eel phase the order parameter, the N\'eel vector, $\vec{S}=\langle \hat{\vec{S}}\rangle$ 
is linear in the spin operator, the VBS long-range order is based on a bilinear scalar combination of $\hat{\vec{S}}$. 
Since broken symmetries in these phases are different, according to the standard Ginzburg-Landau-Wilson paradigm 
a single phase transition between them must be discontinuous. (The actual discontinuities, however, should be very weak if one of the order parameters is characterized by a significant numerical smallness far away from the transition, as is the case in the VBS phase of the \JQ model where the dimer order $\sqrt{D^2} \approx 1/20$  and  no signatures of $Z_4$ broken symmetry are observed even for largest system sizes \cite{Sandvik2007}). Thus, if a single continuous transition
were observed, this would be a strong evidence supporting the second-order DCP scenario. 

The DCP is described by the 3D classical two-component SU(2) symmetric electrodynamics with the emerging 
U(1) gauge vector-field $\vec{A}$ \cite{dcp,dcp1,Motrunich},
$H_{\rm DCP} = \int d^3x \{ t | [\vec{\nabla} - \i \vec{A}] \psi|^2 +\frac{1}{8g} (\vec{\nabla} \times \vec{A})^2 \}$,
where the spinor $\psi$ consists of two complex fields $\psi = (\psi_1,  \psi_2) $. 
According to the mapping, the N\'eel vector $\vec{S}=\frac{1}{2} \vec{n}$ , where $\vec{n}$ is given by 
\beq
\vec{n}= \psi^* \vec{\sigma} \psi,
\label{S}
\eeq
with $\vec{\sigma}$ standing for the Pauli matrices. 
With the noncompact CP$^1$ fixed-modulus constraint \cite{Motrunich}, $|\psi_1|^2 + |\psi_2|^2=1$, one obtains 
$\vec{n}^2= (|\psi_1|^2 + |\psi_2|^2)^2=1$ and $n^+= n_x + in_y = 2 \psi_1 ^* \psi_2$ 
implying that the azimuthal angle of $\vec{n}$ is the relative phase of the
spinon fields, $\varphi= \varphi_2 - \varphi_1$, where $\psi_a \sim \exp(\i\varphi_a),\, a=1,2$.

The lattice version of the DCP action on a simple cubic
lattice \cite{dcp,dcp1} is:
\begin{eqnarray}
H_{\rm DCP}&=&  - t \sum_{\langle ij \rangle,\,  a} \, \left (\psi^*_{ai}\psi^{\:}_{aj}e^{i A_{\langle ij \rangle}}+\mbox{c.c.} \right)+ \nonumber \\
& +& \frac{1}{8g} \sum_{\cal P} \, \left (\vec{\nabla} \times  \vec{A}\right)^2 ,
 \label{DCP}
\end{eqnarray}
where the gauge field $A_{\langle ij \rangle}$ is oriented along the bond $\langle ij \rangle$ from site 
$j$ to site $i$, and $\vec{\nabla} \times \vec{A}$ is the lattice curl operator evaluated on 
elementary plaquettes $\cal P$. The effective constants $(t,g)$ relate in some way to the parameters of 
the \JQ model (\ref{JQ}). Below we will present evidence that $g= 1.1$ and $t=0.8822(4)$ provide the 
closest description of the \JQ model with $J/Q\approx 0.04$ up to a linear size $L\sim L^*=75$.

\noindent {\it Dual variables}. In Ref.~\cite{dejavu}, the statistics of the model (\ref{DCP}) have been 
reformulated in terms of the dual variables---integer bond currents $\vec{J}^{(a)}$ which obey the 
Kirchhoff conservation laws. Accordingly, the partition function of the DCP action $ H_{\rm DCP}$ (\ref{DCP})
can be represented as
\beq
Z= \int d\vec{A}_0 \sum_{\vec{W}_1,\vec{W}_2} Z(\vec{W}_1,\vec{W}_2)\cdot \qquad \qquad
\nonumber \\
\exp  \left[\i \left(\vec{\delta \varphi_1}  + \vec{A}_0\right)\cdot  \vec{W}_1 +
\i \left(\vec{\delta \varphi_2}  + \vec{A}_0\right) \cdot \vec{W}_2 \right], 
\label{ZDCP}
\eeq
where $\vec{A}_0$ stands for the $q=0$ harmonic of the gauge field defined on the lattice with periodic boundary conditions, 
$ \vec{W}_a$ are windings of the bond currents $\vec{J}^{(a)}$, and $\vec{\delta \varphi_a}$ stand for the Thouless boundary phase twists of the spinon-field phases $\varphi_a$. By definition, $ Z(\vec{W}_1,\vec{W}_2)$ is the partition function in a given winding number sector. The integration over $\vec{A}_0$ yields the constraint $ \vec{W}_1 + \vec{W}_2=0$ so that
$Z\, = \, \sum_{\vec{W}}\,  Z(\vec{W}, - \vec{W})\, \exp(\i \, \vec{\delta \varphi} \cdot \vec{W})$ with
$\vec{\delta \varphi}\equiv \vec{\delta \varphi}_1 - \vec{\delta \varphi}_2$.

The stiffness of the S-vector field is found from
\beq
\rho_S \, =\,   \left. \frac{1}{3L} \, \frac{d^2 \ln Z  }{d (\vec{\delta \varphi})^2} \right|_{\vec{\delta \varphi} = 0}=\, 
\frac{1}{3L}  \langle  \vec{W}^2\rangle,
\label{rho_S}
\eeq
It is important that at the critical point the scaling behavior of winding numbers is characterized 
by $\langle  \vec{W}^2\rangle ={\cal O}(1) $ so that $ \rho_S \propto 1/L$. In the ordered N\'eel phase 
$\langle  \vec{W}^2\rangle \propto L$ and the stiffness is finite, $ \rho_S ={\cal O}(1)$.

Our simulations of the \JQ model (\ref{JQ}) are based on the path-integral representation for 
the partition function with periodic boundary conditions in the imaginary time $ 0< \tau \leq \beta$, where 
$\beta$ denotes the inverse temperature (in both cases we employ the worm algorithm approach \cite{WA}, and
simulations of the DCP action were performed as described in Ref.~\cite{dejavu}). 
Accordingly, the spin stiffness $\rho_{JQ}$ with respect to 
the Thouless phase twist can be expressed in terms of the spin worldline windings $W'_x, W'_y$ along 
the spatial directions $x$ and $y$ , respectively:
\beq
\rho_{JQ}= \frac{1}{2\beta} \left[ \langle (W'_x)^2 \rangle + \langle (W'_y)^2 \rangle \right] .
\label{XY}
\eeq

In order to compare the two models at the transition point we also need to fine-tune the $\beta /L$ ratio
for each system size $L$ in order to reach the space-time symmetry in the \JQ model.  We achieve this by defining a space-time symmetric winding in the time direction, $W'_\tau \equiv \sum_{x,y} S_z$ (in the basis where $\hat{S}_z=S_z= \pm 1/2$ is diagonal), and requiring that its mean-square fluctuations coincide with $ \langle (W'_x)^2 \rangle=  \langle (W'_y)^2 \rangle$. We note that  $W'_\tau$ is defined without the factor of 2 (cf.  Eq.(4) of Ref.\cite{Sandvik2010}). Such definition guarantees that  
 fluctuations of $ W'_\tau$ proceed in the same way as the  spatial windings do---in increments of $\pm 1$.

Thus, if parameters of both models (\ref{JQ}) and (\ref{DCP}) are kept
at the critical point $J/Q \approx 0.04$ \cite{Sandvik2007} and $t=t(g)$
(below the bicritical point) \cite{dejavu}, the universal values of the winding number fluctuations
in both models $R_{JQ}= \langle [(W'_x)^2 + (W'_y)^2+ (W'_\tau)^2 ] \rangle  \sim {\cal O}(1) $
and $ R= \langle [(W_x)^2 + (W_y)^2+ (W_z)^2  ]\rangle \sim{\cal  O}(1)$ must coincide provided 
\JQ and noncompact CP$^1$ models have the same fixed point.

\begin{figure}
\vspace*{-0.5cm}
 \includegraphics[width=1.0 \columnwidth]{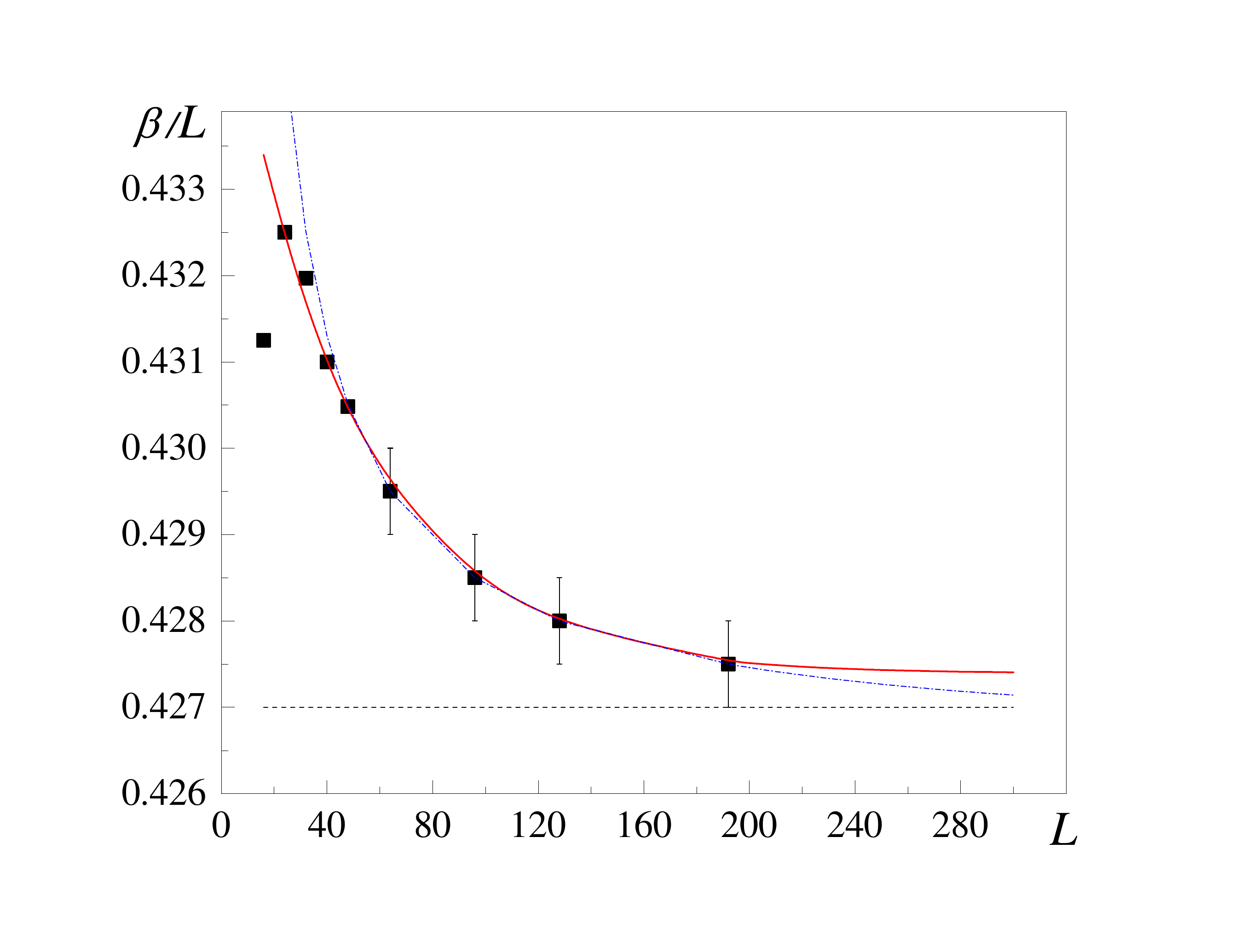}
\vskip-8mm
\caption{(Color online) Optimal ratio $\beta(L)/L$ versus $L$, with the numerical data represented by dots obtained at the pseudocritical points defined in the text. Solid red line is the fit
by $D+ A\exp(-B L)$ and the dash-dotted blue line is the fit by $D + B/L$, with the dashed black line representing the
asymptote $\beta/L =D = 0.4270 \pm 0.0005$ corresponding to the space-time symmetry of the \JQ model.}
\label{beta}
\vskip-5mm
\end{figure}

\begin{figure}
\includegraphics[width=1.0 \columnwidth]{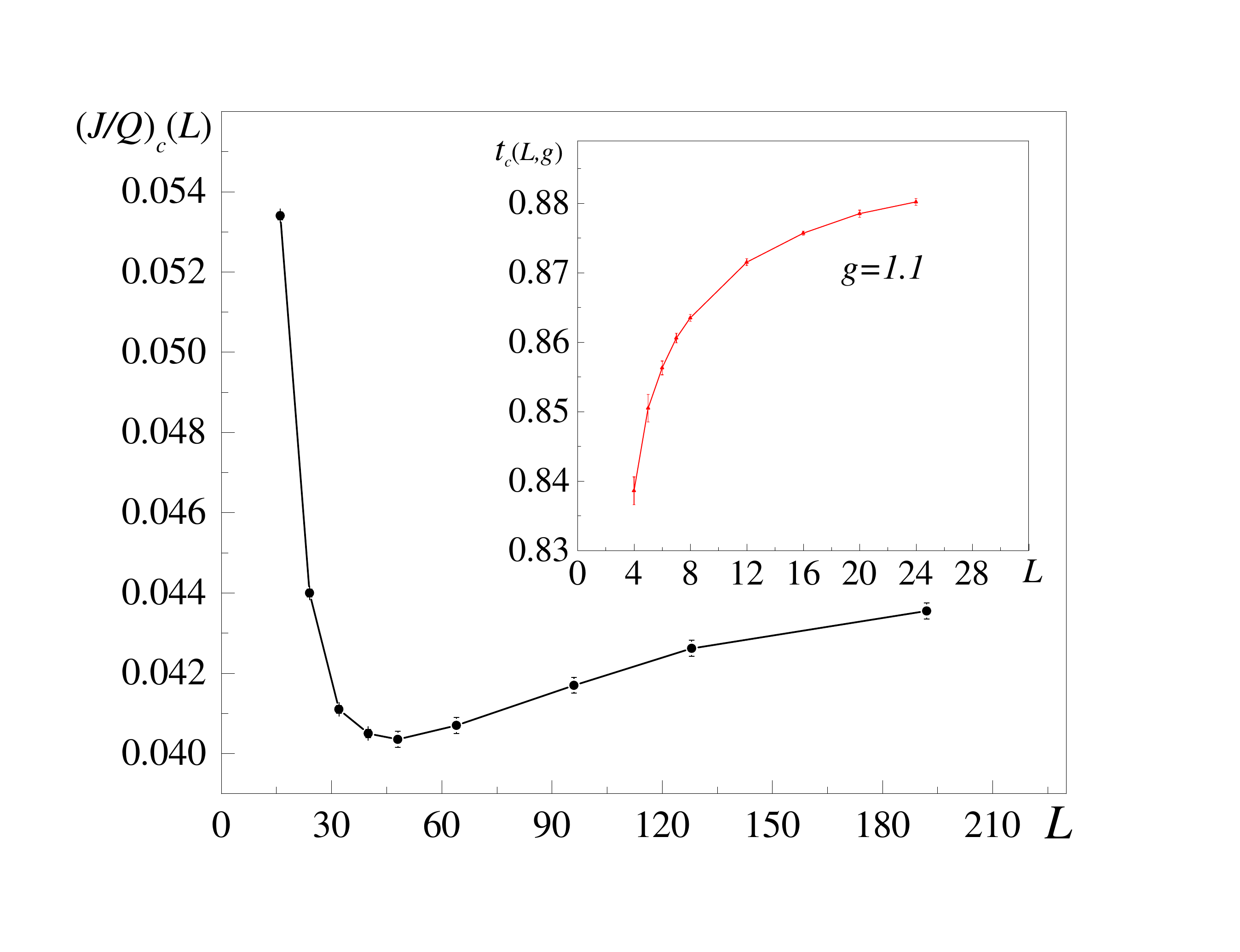}
\vskip-8mm
\caption{(Color online) Size-dependent transition points $(J/Q)_c(L)$ of the \JQ model with the $\beta/L$ ratios 
as in Fig.~\ref{beta}. The inset shows the pseudotransition points $t_c(L,g)$ for $g=1.1$ in the DCP model (\ref{DCP}). 
Extrapolation of both curves to the $L \to \infty$ limit provides estimates of the thermodynamic transition points:
$(J/Q)_c= 0.0451 \pm 0.0004$ and $t_c(g=1.1)= 0.8822 \pm 0.0004$. }
\label{tvsL}
\vskip-5mm
\end{figure}

\noindent {\it Finite size analysis}. Simulations of both models have been conducted
for a sequence of linear sizes using exactly the same definition of the pseudotransition point in a finite size system,
according to the flowgram method \cite{flowgram,dejavu}. Specifically, we tuned model parameters so 
that the ratio of statistical weights of configurations with and without windings, $\cal F$, 
equals the same constant of order unity. We have chosen ${\cal F}=0.55$ because it offers the 
smallest deviations from the space-time symmetry in the \JQ model at large $L$, as shown in Fig.~\ref{beta}.
The values of the parameters at the pseudotransition points for both models are presented in Fig.~\ref{tvsL}.

\begin{figure}
 \includegraphics[width=1.0 \columnwidth]{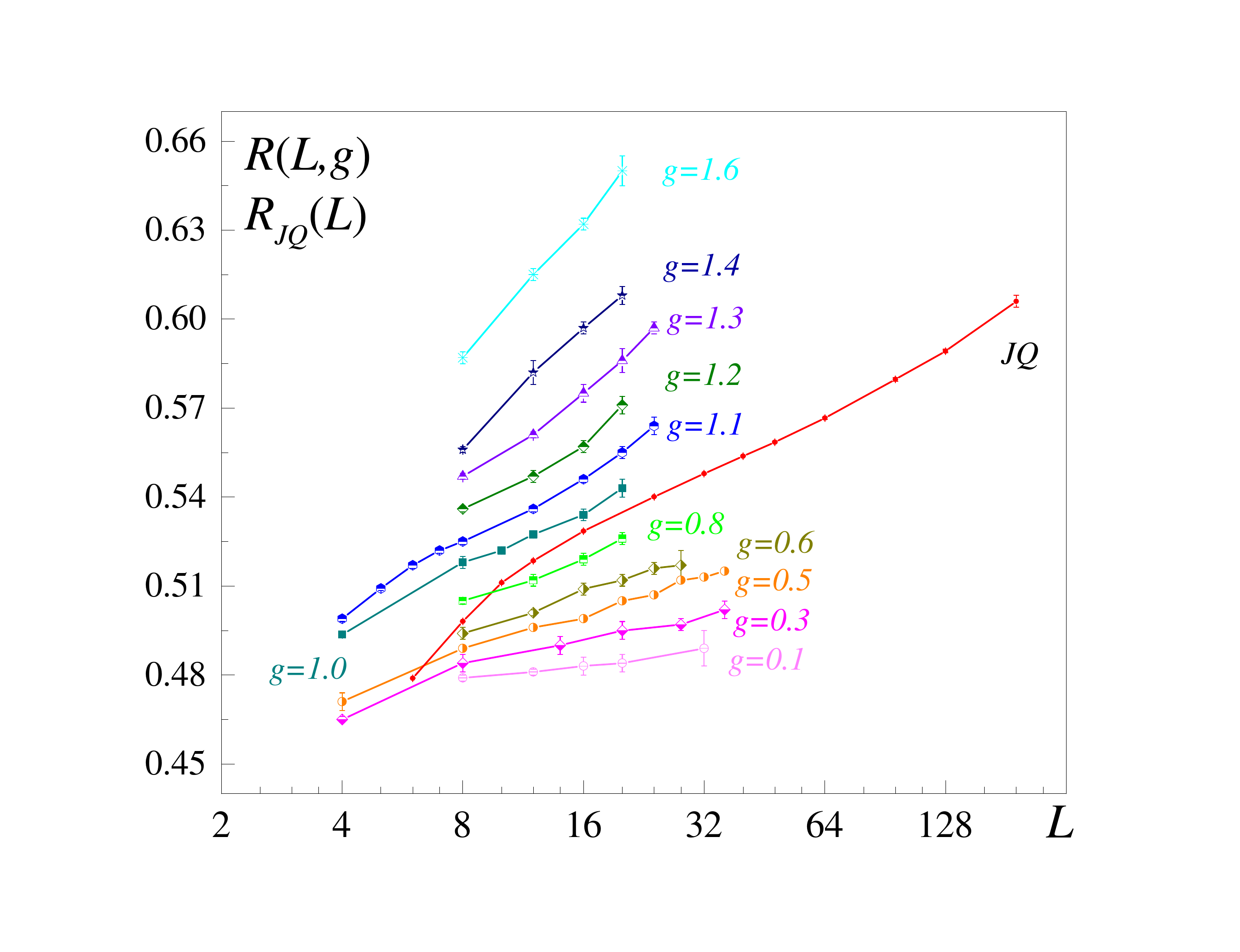}   
\vskip-8mm
\caption{(Color online) Flowgrams of the \JQ (red line) and the DCP models (for several  values of $g$).
}
\label{fig1}
\vskip-5mm
\end{figure}

The universality of scaling behavior is characterized by a unique function 
$ R={\cal R}({\cal F})$ in the thermodynamic limit $L\to \infty,\, \beta \sim L$, i.e. 
for fixed ${\cal F}=0.55$ one expects that $R(L)$ curves saturate to the same value even if they
deviate from each other at finite $L$.
To see if this is indeed the case we have measured $R_{JQ}$ versus $L$ and $R$ versus $(L,g)$
for several  values of $L$ (from $L=4$ to $L=36$ for the DCP model and from $L=6$ to $ L=196$ for the \JQ model).
Figure~\ref{fig1} shows the family of DCP flowgrams $ R(L)$ for several values of the interaction constant $g$. 
It also shows the flowgram $ R_{JQ}(L)$ for the \JQ model. 
It is immediately clear   that the values of $R$-curves overlap and all by itself this is an evidence that DCP theory
captures the physics of the transition point in the \JQ model. This crucial aspect as well as that all the curves feature divergence with $L$, in violation of the scale invariance hypothesis for both models, will become more evident below.
\begin{figure}
 \includegraphics[width=1.0 \columnwidth]{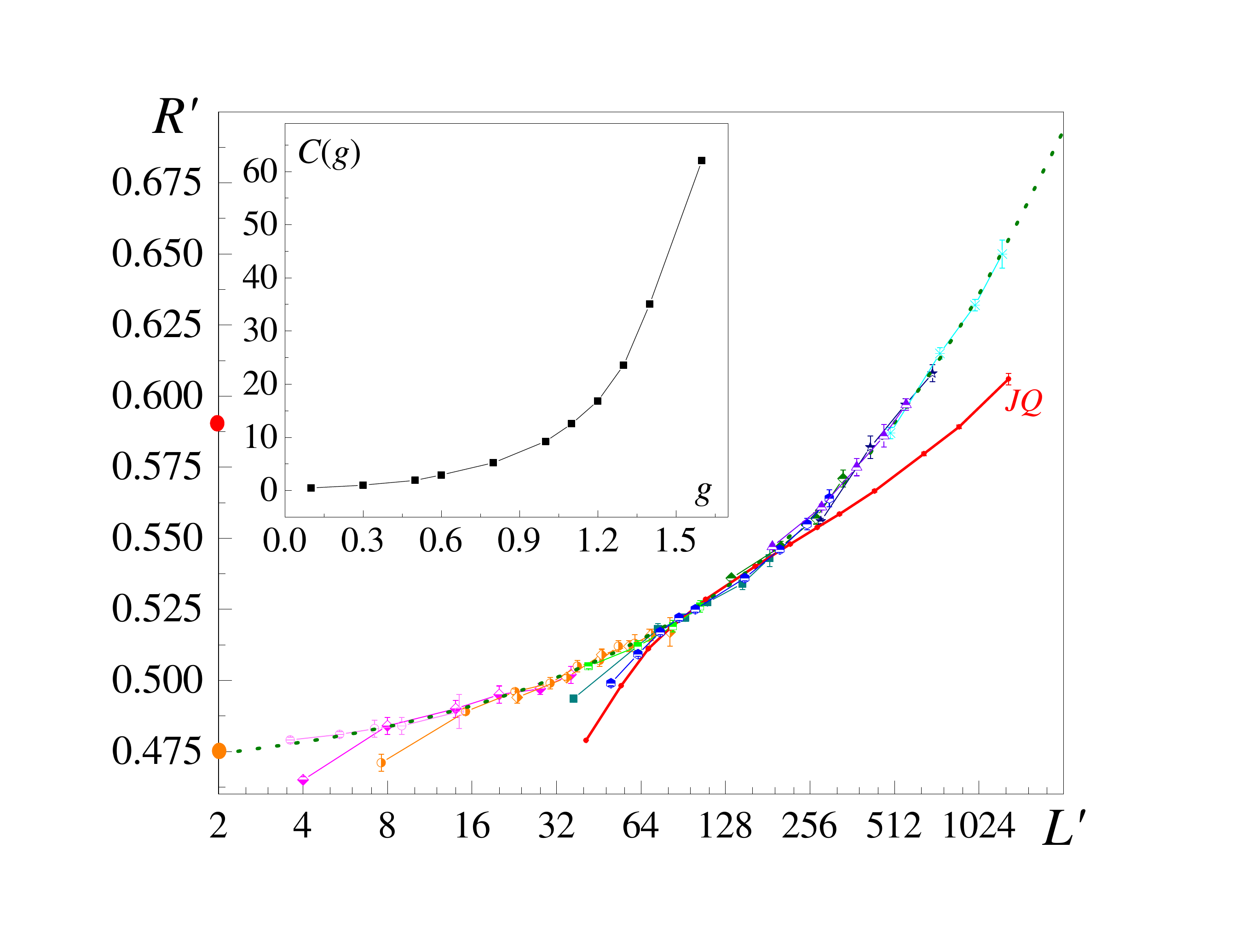}   
\vskip-8mm
\caption{(Color online) Flowgrams from Fig.~\ref{fig1} are collapsed by rescaling system sizes as $L'=C(g)L$ for the DCP model (this amounts to the horizontal shifts of the curves) and $L'=6.8 L$ for the \JQ model. 
Green dotted line shows the master curve fit by the $A+B( L') ^\alpha$ function with  $A=0.463, B=0.00823, \alpha=0.437$. 
The lower (orange) dot on the $R'$-axis indicates the universal value $R'_{\rm O(4)}\approx 0.475$ for the O(4) 
universality class ($g=0$ case). The upper (red) dot on the $R'$ axis corresponds to the universal value  $R'_{\rm O(3)}\approx 0.583$ characterizing the O(3)-universality. Inset: The rescaling function $C(g)$ such that $C(0.3)=1$.
}
\label{fig2}
\vskip-5mm
\end{figure}
\begin{figure}
 \includegraphics[width=1.0 \columnwidth]{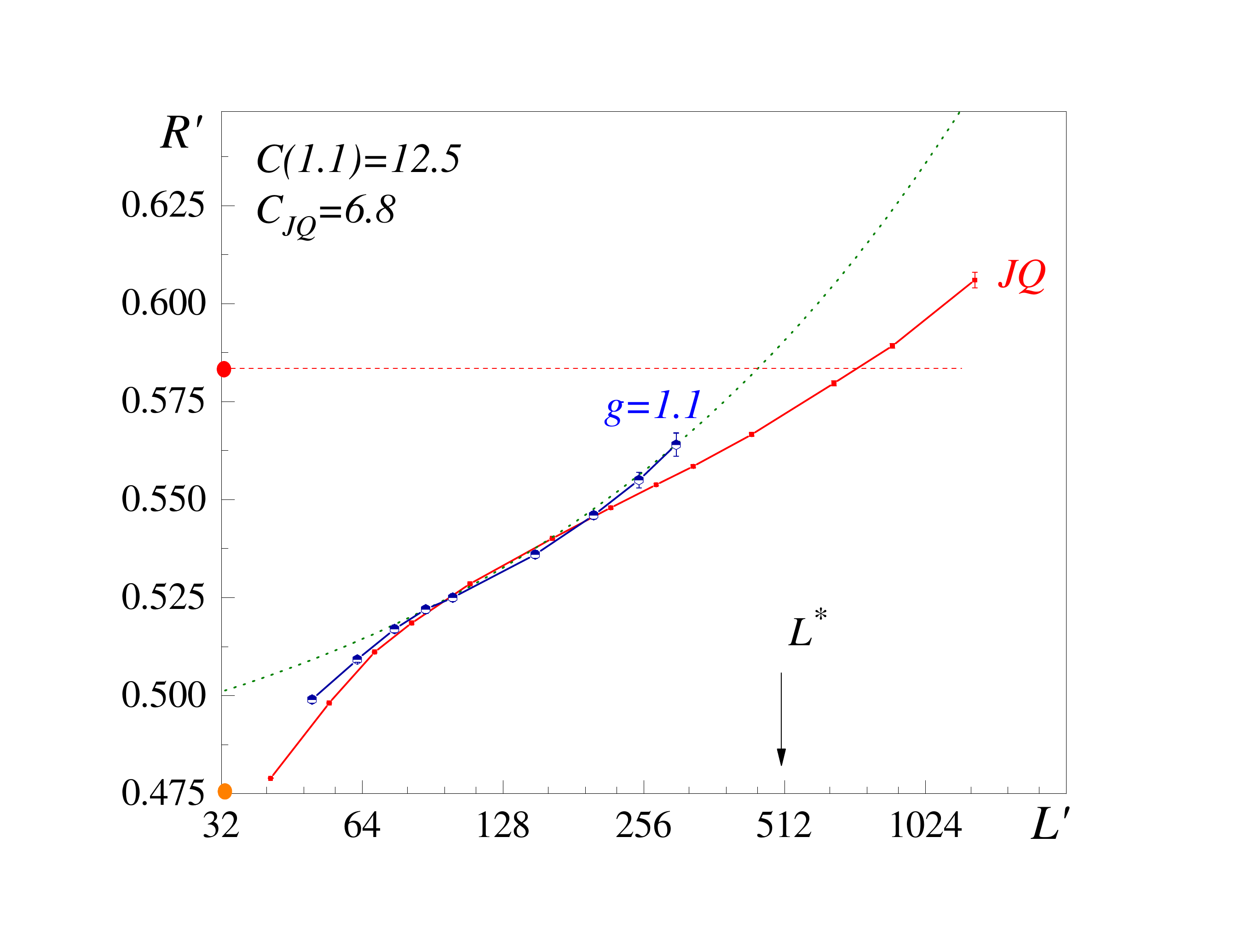}
\caption{(Color online) \JQ flowgram from Fig.~\ref{fig2} is shown together with the DCP $g=1.1$ flowgram demonstrating 
the best overlap between the two models. The dotted line shows the master curve.
The vertical arrow indicates the scale ($L^*=75$ for the \JQ model ) starting from which the flows diverge significantly.  
The dots on the $R'$ axis mark the O(4) and O(3) universal values as in Fig.~\ref{fig2}, 
with the dashed horizontal line for the O(3) asymptote.}
\label{fig3}
\vskip-5mm
\end{figure}

As discussed earlier in Ref.~\cite{dejavu}, the family of DCP flowgrams can be collapsed on a single master curve by  
rescaling system sizes as $L \to C(g)L$, where $C(g)$ is found as a variational distance scale for each value of $g$. 
This collapse implies that properties of the DCP model at coupling strength $g=g_1$ and length scale $L=L_1$ 
are essentially the same as at $g=g_2$ and $L=L_2 = L_1 C(g_1)/C(g_2)$, provided $L$ is larger than some microscopic 
size $\approx 6$. Figure~\ref{fig2} shows the quality of the data collapse procedure as well as the master curve 
which emerges from it. It also shows the flowgram of the \JQ model with rescaled distance $L\to C_{JQ} L$. 
The value of $C_{JQ}$ has been adjusted in order to achieve the best overlap with the DCP-master curve. 
Note that the freedom of choosing $C_{JQ}$ is equivalent to shifting the $R_{JQ}$ curve horizontally 
as a whole (in the $\log L$-scale), i.e. the curve's shape remains preserved. It means that the rescaling procedure is not supposed to 
result in the same slope at the crossing point between the two flows unless they have some common origin.
As can be seen, the two curves coincide with each other at length scales $10<L<50$ (in terms of ``bare" \JQ model sizes) 
before they start significantly diverging from each other at $L \gtrsim L^*=75$.
It is also important that the \JQ flow starts from the O(4) universal value  $R'_{\rm O(4)}\approx 0.475$ rather than from the O(3) universality characterized by $R'_{\rm O(3)}\approx 0.583$ as one would expect from the classical Heisenberg model,
see Fig.~\ref{fig3}. Finally, as Fig.~\ref{fig3} clearly shows, the \JQ flow runs past the O(3) universality at $L>L^*$.

\noindent {\it Conclusion and discussion}.  Our key finding is that the physics of the transition point between the 
N\'eel and insulating VBS phases in the \JQ model is indeed captured by the DCP model up to a large length scale $L^*=75$. 
At small sizes the flows of $R$ and $R_{JQ}$ start from the universal value characterizing the O(4) universality class 
$R'_{\rm O(4)}\approx 0.475$. This very fact is a strong indication that spinons emerge as dominant degrees of freedom
in the \JQ model already at length scales $L<8$ (in agreement with the observed U(1) symmetry of the VBS order 
parameters \cite{Sandvik2007}). However, the divergence of the flows at $L>L^*$ unambiguously excludes the 
possibility that the \JQ model and the DCP action share the same criticality in the thermodynamical limit.

As shown in Ref.~\cite{dejavu}, the runaway flow of the DCP master curve ends up in the first-order phase transition 
(detectable at $g\approx 1.65$ for sizes $L \sim 30 -36$). [The rescaling function $C(g)$ shown in the inset in 
Fig.~\ref{fig2} is a smooth function defined on $g\geq 0$. It has no features 
indicating the presence of the tricritical point at some $g=g_{\rm tr}>0$].
This explains why the \JQ and DCP flows ultimately depart. Given the data, there are two possibilities
for the ultimate fate of the \JQ flow: either the first-order transition or some unknown universality at 
larger values of $R'$. 
The fact that both models follow the same flow at $L<L^*$ and both violate the scale invariance hypothesis 
at large length scales strongly favors the first possibility---while showing quasiuniversal behavior 
at intermediate $L$ the two models deviate from this universality when the system size is 
approaching the size of the first-order nucleation bubble which does not need to be the same in 
different models. 

A. B. K acknowledges helpful discussions of the numerical aspects of the DCP model with Matthias Troyer and of the main result with Leon Balents. We also acknowledge useful discussions of the J-Q model with Anders Sandvik. This work was supported by the National Science Foundation under Grants No. PHY1005527, No. PHY1005543,
and by a grant for computer time from the CUNY HPCC under NSF Grants No. CNS-0855217 and No. CNS - 0958379.
KC, YH and YD acknowledge support from NNSFC under Grant No. 11275185, CAS,
and NKBRSFC under Grant No. 2011CB921300.

\end{document}